\documentclass[twocolumn,english,prl,floatfix,citeautoscript,nofootinbib]{revtex4}
\usepackage{amsfonts}
\usepackage{amsbsy}
\usepackage{latexsym,epsfig,graphicx}
\usepackage{dcolumn}
\usepackage{subfigure}
\usepackage{comment}
\usepackage{color}
\usepackage[colorlinks,urlcolor=blue,citecolor=blue]{hyperref}
\usepackage{amstext}
\usepackage{amssymb}
\usepackage{setspace}
\usepackage{amsmath}
\usepackage{amssymb}
\usepackage{bm}

\input{tcilatex}
\begin{document}

\title{Higher-order Topological Phases of Magnons in van der Waals Honeycomb Ferromagnets}
\author{Yun-Mei Li}
\email{yunmeili@xmu.edu.cn}
\affiliation{Department of Physics, School of Physical Science and Technology, Xiamen University, Xiamen 361005, China}
\author{Ya-Jie Wu}
\affiliation{School of Science, Xi'an Technological University, Xi'an 710032, China}
\author{Xi-Wang Luo}
\affiliation{CAS Key Laboratory of Quantum Information, University of Science and Technology of China, Hefei 230026, China}
\author{Yongwei Huang}
\affiliation{School of Physics and Electronic-Electrical Engineering, Ningxia University, Yinchuan 750021, China}
\author{Kai Chang}
\affiliation{SKLSM, Institute of Semiconductors, Chinese Academy of Sciences, P.O. Box 912, Beijing 100083, China}

\begin{abstract}
  We theoretically propose a second-order topological magnon insulator by stacking the van der Waals
  honeycomb ferromagnets with antiferromagnetic interlayer coupling.
  The system exhibits $Z_{2}$ topological phase, protected by pseudo-time-reversal symmetry (PTRS).
  An easy-plane anisotropy term breaks PTRS and destroys the topological phase.
  Nevertheless, it respects a magnetic two-fold rotational symmetry which protects a second-order topological phase with corner modes in bilayer and hinge modes along stacking direction. Moreover, an introduced staggered interlayer coupling establishes a $Z_{2}\times Z$ topology, giving rise to gapped topological surface modes carrying non-zero Chern numbers. Consequently, chiral hinge modes propagate along the horizontal hinges
  in a cuboid geometry and are robust against disorders.
  Our work bridges the higher-order topology and magnons in van der Waals platforms,
  and could be used for constructing topological magnonic devices.
\end{abstract}

\maketitle

In recent years, higher-order topological phases of matter have attracted much attention due to the
unconventional bulk-boundary correspondence~\cite{WABenalcazar,ZSong,JLangbehn,WABenalcazar2,FSchindler,MEzawa,XZhu,MGeier,
KKudo,ZYan,XWLuo,YJWu,CChen,YRen,WWang}. Conventional topological insulator (TI)
and superconductor (TSC) have one-dimensionally (1D) lower boundary or surface modes~\cite{MZHasan,XLQi}.
As a contrast, $r$th-order $(r\geq2)$ TIs and TSCs in $d$ dimension host $(d-r)$-dimensional topologically protected modes,
dubbed corner or hinge modes. These higher-order topological phases are first proposed in the electronic systems,
predicted and observed
in quite a few materials~\cite{CChen,CYue,ZWang,YXu2,XLSheng,BLiu,YBChoi,BJack}, and then
extended to the bosonic systems, such as photons~\cite{JNoh,XDChen,BYXie}, phonons~\cite{MSGarcia,XZhang,LLuo},
and the magnons~\cite{ZLi,ASil,THirosawa,ZXLi1,ZXLi2,AMook,MJPark,ZXLi3} in magnetic systems.
For now, the works on higher-order topological phases of magnons are mainly distributed in the magnetic solitonic system.
These works in the spin systems utilize either specially designed coupling~\cite{AMook,ASil}
or externally controlled antiferromagnetic noncollinear order ~\cite{MJPark},
which are not realized experimentally due to technique difficulty.

Since the experimental discovery of two-dimensional (2D) ferromagnetism~\cite{BHuang,CGong,ZFei}, the van der Waals magnets provide new and versatile platforms to investigate the fundamental physics and also new opportunities for spintronic device design.
Recent experimental progresses reveal topological phases in 2D magnetic materials, such as CrI$_{3}$ and MnBi$_{2}$Te$_{4}$, etc~\cite{LChen1,LChen2,FZhu,YDeng,DZhang,CLiu,BWei}.
The van der Waals magnets show a store of in-plane and out-of-plane magnetic orders.
For example, in the collinear phase, there are four representative ground states, the combination
of intralayer ferromagnetic (FM)/antiferromagnetic (AFM) and interlayer FM/AFM orders.
When involving non-ferromagnetic orders, the magnon behavior usually does not have direct counterpart
to electronic and other (quasi)particle systems~\cite{YHLi,CWang,JSklenar,HYYuan,GChen,HKondo,YMLi}.
The interlayer magnetic orders and coupling strength can be even tuned by the external field, pressure, stacking and the interlayer distance~\cite{PJiang,XKong,TLi,TSong}.
As a result, the van der Waals magnets offer us more degrees of freedom to seek
for intriguing properties of magnons,
with experimental feasibility by use of the state-of-the-art van der Waals engineering.

In this Letter, motivated by the AFM interlayer coupling in chromium trihalides, we theoretically proposed that AA-stacked honeycomb ferromagnets with AFM interlayer coupling host second-order topological phases for magnons. The monolayer is a Chern insulator in the presence of next-nearest-neighboring (NNN) Dzyaloshinskii-Moriya interactions (DMI). The interlayer AFM coupling endows a pseudo-time-reversal symmetry (PTRS), realizes $Z_{2}$ topological phases and
supports helical edge states in bilayer and anisotropic topological surface modes (TSMs) in bulk located on the surfaces parallel to the stacking direction.
An easy-plane anisotropic term breaks the PTRS and also the $Z_{2}$ topological phase. But a magnetic two-fold rotational symmetry is preserved.
Along the lines in the Brillouin zone leaving the system invariant under the rotational symmetry,
the Hamiltonian can be decomposed into two parts with different eigenvalues of the symmetry operator.
A half-quantized rotation-graded topological polarizations are obtained, indicating the system hosts a second-order topological phase with corner modes in bilayer and hinge modes along the stacking direction. Besides, a staggered interlayer coupling builds a $Z_{2}\times Z$ topology,
giving rise to gapped TSMs located on the surfaces perpendicular to the stacking direction, carrying non-zero Chern numbers.
According to the bulk-boundary correspondence, chiral hinge modes propagate along the horizontal hinges with
opposite chirality on the top and bottom horizontal hinges in a cuboid geometry.
We would like to emphasize that our conclusion can be applied to different stacking, not limited to the AA stacking.
This makes our proposal easier to be realized experimentally.

\begin{figure}[t]
  \centering
  \includegraphics[width=0.45\textwidth]{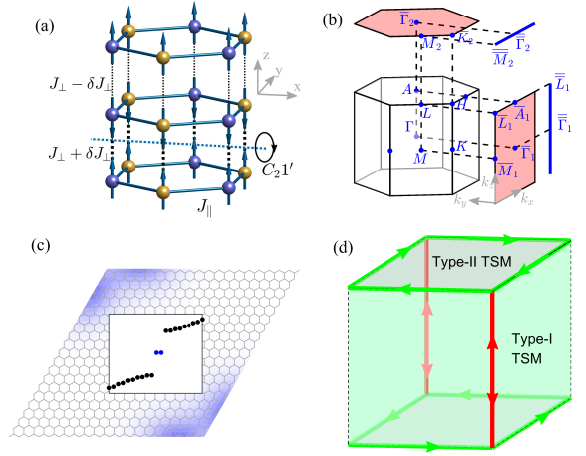}\\
  \caption{(a) The unit cell of AA stacked antiferromagnetic coupled honeycomb ferromagnets. The AFM interlayer coupling is staggered.
  The horizontal dotted line indicates a magnetic two-fold rotation axis (grey group).
   (b) The corresponding Brillouin zone (BZ) and projected 2D and 1D BZ. (c) The corner modes in bilayer flake.
   The color represents the local density distribution of corner modes. The inset is the magnon eigenvalues of the flake geometry
   with $J_{\perp}=0.3J_{\parallel}$, $K_{z}=0.2J_{\parallel}$, $K_{x}=J_{\parallel}$, $D=0.05J_{\parallel}$.
   (d) Illustration of Topological surface and hinge modes. The blue lines denote the achiral hinge modes while the green ones denote chiral hinge modes. }\label{fig1}
\end{figure}

\begin{figure}[t]
  \centering
  \includegraphics[width=0.5\textwidth]{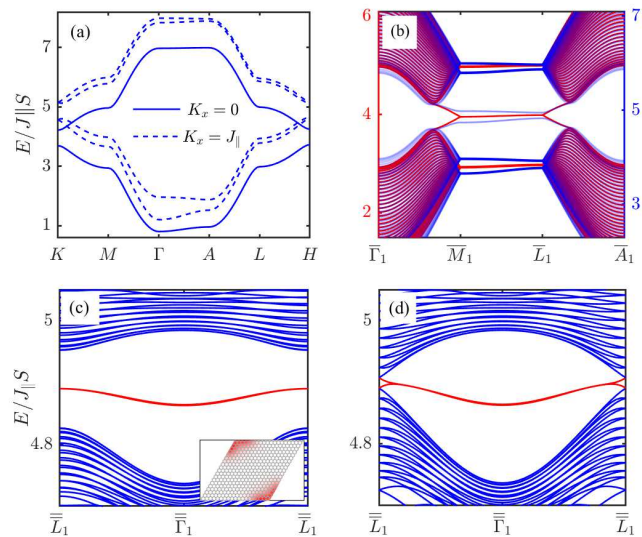}\\
  \caption{(a). The bulk magnon band structure at $K_{x}=0$ (solid) and $K_{x}=J_{\parallel}$ (dashed).
   In both, $K_{z}=0.2J_{\parallel}$, $D=0.05J_{\parallel}$, $J_{\perp}=0.3J_{\parallel}$, $\delta J_{\perp}=0.15J_{\parallel}$.
   In subsequent calculation, we adopt the same parameters when not pointing out in particular.
   At nonzero $K_{x}$, the degeneracy of magnon bands are lifted.
   (b) The magnon band in the $k_{x}-k_{z}$ plane with a finite unit cell number $N_{y}=20$ along y direction
   at $K_{x}=0$ (red) and $K_{x}=J_{\parallel}$ (blue). In all the calculation of finite
  size magnon spectrum, we adopt a compensated boundary.
  (c). Magnon bands along $k_{z}$ axis with stacked infinite parallelogram flake. The red lines are the hinge modes with doubly degeneracy.
  The inset shows the local density distribution of the hinge modes at $k_{z}=\pi/2$.  (d). Magnon bands same in (c) but  with $\delta J_{\perp}=0$.
  }\label{fig2}
\end{figure}

\textit{Model--}We consider the AA-stacked van der Waals honeycomb ferromagnets for simplicity, as shown in Fig.~\ref{fig1} (a).
The interlayer coupling is AFM-type, and the strength is staggered controlled by the interlayer distance.
The minimal spin interaction Hamiltonian is given by
\begin{equation}\label{eq1}
  H=\sum_{l}H_{l}+\sum_{i,l}[J_{\perp}+(-1)^{l+1}\delta J_{\perp}]\mathbf{S}_{i}^{l}\cdot\mathbf{S}_{i}^{l+1},
\end{equation}
where $l$ is the layer index. $J_{\perp}\pm \delta J_{\perp}>0$ describe the AFM coupling and the intralayer spin Hamiltonian
\begin{eqnarray}
  H_{l}=&-&J_{\parallel}\sum_{\langle ij\rangle}\mathbf{S}_{i}^{l}\cdot\mathbf{S}_{j}^{l}
  +D\sum_{\langle\langle ij\rangle\rangle}\nu_{ij}\hat{z}\cdot(\mathbf{S}_{i}^{l}\times\mathbf{S}_{j}^{l}) \nonumber \\
  &+&\sum_{i}[K_{x}(S_{i,x}^{l})^2-K_{z}(S_{i,z}^{l})^2]. \label{eq2}
\end{eqnarray}
$J_{\parallel}>0$ characterizes FM intralayer coupling and $D$ represents the NNN DMI strength. $\nu_{ij}=\pm 1$
with $+$ ($-$) for counterclockwise (clockwise) circulation.
The last two terms denote the single-ion easy-plane (EPA) and easy-axis anisotropy.
There are two ground states of the spin configuration, the down layer pointing up while the upper layer pointing down, or reversed.
For convenience, we choose one of the two spin configurations without loss of generality.
By applying the Holstein-Primakoff transformation $S_{+}=\sqrt{2S}a$, $S_{-}=\sqrt{2S}a^{\dagger}$, $S_{z}=S-a^{\dagger}a$ for first layer
and $S_{-}=\sqrt{2S}a$, $S_{+}=\sqrt{2S}a^{\dagger}$, $S_{z}=-S+a^{\dagger}a$ for second layer, and making the Fourier transformation
$a_{i}^{l}=\frac{1}{N}\sum_{\mathbf{k}}e^{i\mathbf{k}\cdot\mathbf{r}_{i}}b_{\alpha,\mathbf{k}}^{l}$,
$a_{i}^{l\dagger}=\frac{1}{N}\sum_{\mathbf{k}}e^{-i\mathbf{k}\cdot\mathbf{r}_{i}}b_{\alpha,\mathbf{k}}^{l\dagger}$ ($\alpha=A,B$, the sublattice index),
the Hamiltonian in $\mathbf{k}$-space is expressed as $H=\sum_{\mathbf{k}}\Psi_{\mathbf{k}}^{\dagger}H_{\mathbf{k}}\Psi_{\mathbf{k}}$ with
bosonic Bogoliubov-de Gennes (BdG) Hamiltonian in $\mathbf{k}$-space~\cite{SM}
\begin{equation}\label{eq3}
  H_{\mathbf{k}}=\left(\begin{array}{cccc}
  h_{1}(\mathbf{k}) & 0 & K_{x} & h_{\perp,k_{z}} \\
  0 & h_{2}(\mathbf{k}) & h_{\perp,k_{z}} & K_{x} \\
  K_{x} & h_{\perp,k_{z}}^{\dagger} & h_{1}^{T}(-\mathbf{k}) & 0 \\
  h_{\perp,k_{z}}^{\dagger} & K_{x} & 0 &  h_{2}^{T}(-\mathbf{k})
  \end{array}\right).
\end{equation}
The basis is $\Psi_{\mathbf{k}}=(\psi_{\mathbf{k}},\psi_{-\mathbf{k}}^{\dagger})^{T}$ with $\psi_{k}=(b_{A,\mathbf{k}}^{1},b_{B,\mathbf{k}}^{1},b_{A,\mathbf{k}}^{2},b_{B,\mathbf{k}}^{2})^{T}$.
$h_{1}(\mathbf{k})=h_{0}\sigma_{0}+\sum_{i=x,y,z}h_{i}\sigma_{i}$, $h_{2}(\mathbf{k})=h_{1}^{*}(-\mathbf{k})$,
$h_{\perp,k_{z}}=(J_{\perp}+\delta J_{\perp})S+(J_{\perp}-\delta J_{\perp})Se^{-ik_{z}}$.
$\sigma_{0}$ is $2\times2$ identity matrix and $\sigma_{i=x,y,z}$ the Pauli matrices.
$h_{0}=(2K_{z}+K_{x}+3J_{\parallel}+2J_{\perp})S$, $h_{x}=Re(\gamma_{\mathbf{k}})$, $h_{y}=-Im(\gamma_{\mathbf{k}})$,
$h_{z}=-2DS\sum_{i=1}^{3}\sin(\mathbf{k}\cdot\mathbf{a}_{i})$, $\gamma_{\mathbf{k}}=-J_{\parallel}S (1+e^{-i\mathbf{k}\cdot\mathbf{a}_{3}}+e^{i\mathbf{k}\cdot\mathbf{a}_{2}})$,
$\mathbf{a}_{1}=(\sqrt{3},0,0)$, $\mathbf{a}_{2}=(-\frac{\sqrt{3}}{2},\frac{3}{2},0)$, $\mathbf{a}_{3}=(-\frac{\sqrt{3}}{2},-\frac{3}{2},0)$ intralyer NNN vectors. Diagonalizing the quadratic form in Eq. (\ref{eq3}) invokes the Bogoliubov transformation,
$T_{\mathbf{k}}^{\dagger}H(\mathbf{k})T_{\mathbf{k}}=diag\{E_{\mathbf{k}},E_{-\mathbf{k}}\}$, with paraunitary eigenvectors satisfying
$T_{\mathbf{k}}^{\dagger}\tau_{z}T_{\mathbf{k}}=\tau_{z}$ and $\tau_{z}$ is the Pauli matrix acting on the particle-hole space~\cite{RShindou}.
Typical magnon bands are shown in Fig.~\ref{fig2} (a).

In the absence of EPA term, the Hamiltonian in Eq.~(\ref{eq3}) hosts a PTRS,
$\Theta\tau_{z}H_{\mathbf{k}}=\tau_{z}H_{-\mathbf{k}}\Theta$ with $\Theta=i\tau_{z}s_{y}\sigma_{0}\mathcal{K}$, where
$s_{y}$ is Pauli matrix acting on the layer index and $\mathcal{K}$ denotes the complex conjugation.
We can see this PTRS coming from the AFM interlayer coupling, as the intralayer ferromagnetism
breaks the TRS. As a result, the magnon bands show Kramers degeneracy (Fig.~\ref{fig2} (a)), and also
host nontrivial Z$_{2}$ topological phase.
The previous work~\cite{LFu} on quantum spin Hall phase and topological crystalline insulators told us
four Z$_{2}$ invariants were needed to form the combination
$[\nu_{0},(\nu_{1},\nu_{2},\nu_{3})]$ to describe the topological phase in 3D.
As the bosonic Hamiltonian (\ref{eq3}) does not host an inversion symmetry. We calculate the Z$_{2}$ invariants
by enlarging the unit cell along $y$-direction to get a cuboid BZ~\cite{RYu} and obtain
$\nu_{0}=0$, $(\nu_{1},\nu_{2},\nu_{3})=(001)$, indicating the system
is in a ``weak" TI phase~\cite{SM}.
Consequently, topological surface modes (TSMs) appear on the surfaces parallel to the stacking direction, i.e. the $z$-axis, labelled as type-I.
In Fig.~\ref{fig2} (b), we plot the magnon bands in the projected $k_{x}$-$k_{z}$ plane with finite width along the $y$ axis.
The anisotropic gapless TSMs emerge in the gap and show a nodal line along the $k_{z}$ axis, not even number of surface Dirac cones
exhibited in the previous works~\cite{LFu,YYang,SHKooi}.
The $Z_{2}$ invariant for a bilayer such system
should be equal to $\nu(k_{z}=0)=1$, indicating a 2D $Z_{2}$ TI phase and gapless Dirac dispersive edge states~\cite{SM}.

\textit{Second-order topology--}In the presence of EPA term, the PTRS is no longer kept. The magnon band degeneracy is lifted, as
shown in Fig.~\ref{fig2} (a), thus the $Z_{2}$ topological phase is broken. As a consequence, type-I TSMs at zigzag-terminated surfaces are gapped, as shown
by the blue curves in Fig.~\ref{fig2} (b).
Interestingly, when we consider a parallelogram pillar geometry with two zigzag-terminated surfaces encounter with
an armchair connection, hinge modes arise in the surface gap, as shown by the red curves in Fig.~\ref{fig2} (c), where the magnon dispersion along $k_{z}$
direction is plotted. The local density distribution of the hinge modes is shown in the inset.
Here these hinge modes come from nontrivial bulk band topology.
Although the EPA breaks the PTRS and drives the 3D TI into a trivial insulator, it preserves various crystalline symmetries, i.e.,
a magnetic two-fold rotation symmetry in our case shown in Fig.~\ref{fig1} (a), which make the system to exhibit topological crystalline phase.

The magnetic two-fold rotational symmetric operation, illustrated in Fig.~\ref{fig1} (a), exchanges the
A (B) sublattice in the first layer with the B(A) sublattice in the second layer and flips the spin.
Acting on the magnon Hamiltonian ((\ref{eq3})), we have
$\mathcal{M}_{y}\tau_{z}H_{\mathbf{k}}(k_{x},k_{y},k_{z})\mathcal{M}_{y}=\tau_{z}H_{\mathbf{k}}(k_{x},-k_{y},k_{z})$,
with $\mathcal{M}_{y}=\tau_{0}s_{x}\sigma_{x}$,
behaving like a mirror symmetry changing $y$ to $-y$. As $\mathcal{M}_{y}^{2}=1$, $\mathcal{M}_{y}$ has two eigenvalues
$\pm 1$. This indicates that we can decompose the Hamiltonian ((\ref{eq3})) into decoupled two parts labelled by the
eigenvalues of $\mathcal{M}_{y}$ along the lines $l_{\mathcal{M}_{y}}$ in the BZ that leave the Hamiltonian invariant after applying
$\mathcal{M}_{y}$. Here we regard the $k_{z}$ as a parameter.
There are two types of lines: along $k_{x}$ at $k_{y}=0$ and $2\pi/3$. We mainly discuss the case $k_{y}=0$, which is directly
related to the hinge modes. The Hamiltonian at $k_{y}=0$ can be decomposed into two parts $H_{\mathbf{k}}^{\pm}(k_{y}=0)$ by using the
$\mathcal{M}_{y}$ eigenvectors
\begin{equation}\label{eq4}
  H_{\mathbf{k}}^{\pm}=\left(
  \begin{array}{cccc}
  h_{0}+h_{z}^{0} & h_{x}^{0} & K_{x} & \pm h_{\perp,k_{z}} \\
  h_{x}^{0} & h_{0}-h_{z}^{0} & \pm h_{\perp,k_{z}} & K_{x} \\
  K_{x} & \pm h_{\perp,k_{z}}^{\dagger} & h_{0}-h_{z}^{0} & h_{x}^{0} \\
  \pm h_{\perp,k_{z}}^{\dagger} & K_{x} & h_{x}^{0} & h_{0}+h_{z}^{0}
  \end{array}
  \right),
\end{equation}
where $h_{x}^{0}$ and $h_{z}^{0}$ take the value of $h_{x}$ and $h_{z}$ at $k_{y}=0$ and $\pm$ denotes the subspace with
$\pm 1$ eigenvalues.

We can diagonalize the bosonic Hamiltonians $H_{\mathbf{k}}^{\pm}$ by employing
the Bogoliubov transformation, $T_{\mathbf{k},\pm}^{\dagger}H_{\mathbf{k}}^{\pm}T_{\mathbf{k},\pm}=diag\{E_{\mathbf{k}}^{\pm},E_{-\mathbf{k}}^{\pm}\}$, with paraunitary eigenvectors satisfying $T_{\mathbf{k},\pm}^{\dagger}\tau_{z}T_{\mathbf{k},\pm}=\tau_{z}$.
Then we have the rotation-graded Berry connection $A_{\mu}^{n,\pm}=i\mathrm{Tr}[\Gamma^{n}\tau_{z}T_{\mathbf{k},\pm}^{\dagger}\tau_{z}(\partial_{k_{\mu}}T_{\mathbf{k},\pm})]$ for n-th band, where
$\Gamma^{n}$ is the diagonal matrix taking $+1$ for n-th digaonal component and zero otherwise.
The rotation-graded topological polarization
$P_{\mathcal{M}_{y}}^{\pm}(k_{z})=\frac{1}{2\pi}\sum_{n}\int_{l_{\mathcal{M}_{y}}}\mathbf{A}_{n}^{\pm}\cdot d\mathbf{l}_{\mathcal{M}_{y}}$. The summation is over the occupied particle bands. In our case, the integration interval is $(0, \frac{4\pi}{\sqrt{3}})$ along $k_{x}$ at $k_{y}=0$.
We calculate numerically the topological polarization $P_{\mathcal{M}_{y}}^{\pm}$ for rotation-graded mangnon bands
and find that $P_{\mathcal{M}_{y}}^{\pm}=1/2$ at any $k_{z}$.
This half-quantized polarization confirms that the system is in a topological crystalline phase~\cite{TMLec}. Gapless TSMs would exist on the armchair surface even at nonvanishing EPA, different from the TSMs at zigzag surfaces that are gapped by the EPA.
Besides, when two zigzag surfaces encounter with an armchair connection, in-gap modes will emerge in the surface bandgap, as verified above in Fig.~\ref{fig2} (c).
As the $P_{\mathcal{M}_{y}}^{\pm}$ is independent on $k_{z}$, the hinge modes are not chiral and are nearly flat
due to the weak interlayer coupling.
Thus they are not robust against disorders.
But as long as the surface gap remains, these hinge modes retain.
In Fig.~\ref{fig2} (d), we plot the magnon dispersion along $k_{z}$ at $\delta J_{\perp}=0$.
The surface gap closes at $k_{z}=\pi$, the hinge modes disappear near $k_{z}=\pi$.
The above discussions can also be applied to the bilayer case.
The bilayer is also in second-order topological phase with corner modes~\cite{SM}, as shown in Fig.~\ref{fig1} (c).
These corner modes are also protected by the magnetic rotational symmetry.

We now discuss the topology along the $k_{z}$ direction.
The EPA term breaks the Z$_{2}$ topology of the system and gives topological hinge modes.
Here we show that at nonzero $K_{x}$, there exists another Z$_{2}$ topology along the stacking direction and will generate another type of
TSMs and also chiral hinge modes. We start from the Bosonic BdG Hamiltonian in Eq.~(\ref{eq3}) and introduce the Berry connection
${A}_{\mu}^{n}(\mathbf{k})=i\mathrm{Tr}[\Gamma^{n}\tau_{z}T_{\mathbf{k}}^{\dagger}\tau_{z}(\partial_{k_{\mu}}T_{\mathbf{k}})]$
and the Berry curvature $\bm\Omega^{n}=\nabla_{\mathbf{k}}\times\mathbf{A}^{n}(\mathbf{k})$.
To characterize the bulk polarization along the stacking direction, we compute the
Wannier bands $w_{z}^{n}=(1/{2\pi})\int_{-\pi}^{\pi}\mathbf{A}_{z}^{n}dk_{z}$.
We find when $\delta J_{\perp}<0$,
$w_{z}=1/2$ for all the bands and for all $(k_{x},k_{y})$ at any finite $K_{x}$.
We have checked $w_{z}=0$ for $\delta J_{\perp}>0$. This indicates
a quantized bulk polarization along staking direction $P_{z}=[1/V_{\Omega}]\int_{\Omega} w_{z}dk_{x}dk_{y}=1/2$ for all the bands,
where $\Omega$ is projected ontop BZ shown in Fig.~\ref{fig1} (b) and $V_{\Omega}$ the area.
The above analysis shows an additional $Z_{2}$ invariant in our system, coming from the staggered interlayer coupling, which gives
a Su-Schrieffer-Heeger (SSH) configuration of antiferomagnetic order.

\begin{figure}[t]
  \centering
  \includegraphics[width=0.5\textwidth]{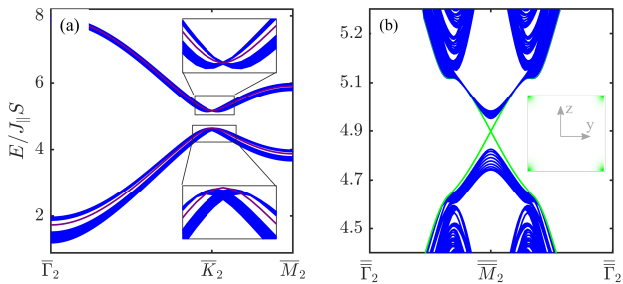}\\
  \caption{(a) Magnon bands with finite FM layers $2N_{z}=30$, $N_{z}$ are the number of unit cell along stacking direction.
  $\delta J_{\perp}=-0.15J_{\parallel}$. The purple lines indicate the type-II TSMs. Each line are double degenerate.
   (b) Magnon bands along $k_{x}$ direction with finite unit cells along $y$ and $z$ direction $N_{y}=20$, $N_{z}=15$.
   The green lines denote the hinge modes. The inset shows the local density distribution of hinge modes in the $yz$ plane.
   }\label{fig3}
\end{figure}

The quantized polarization will give rise to a different type of TSM, labelled by type-II.
In Fig.~\ref{fig3} (a), we plot the magnon bands in $k_{x}-k_{y}$ plane with finite layers along stacking direction.
Gapped TSMs emerge between the split bands, not in the bulk bandgap.
These TSMs are doubly degenerate, distributed on the top and bottom surfaces, respectively.
We calculate the Chern numbers for the type-II TSMs with lower energy and find the Chern number is $+1$ ($-1$) for type-II TSMs located at bottom (top) surface.
Due to the nontrivial Chern numbers, a class of $Z_{2}\times Z$ topology is established even the PTRS is broken.
According to the bulk-boundary correspondence, there exists topological modes closing the
TSM bandgap. In Fig.~\ref{fig3} (b), we further impose a finite width constraint on the $y$-direction,
and verified the existence of gapless states. These states are thus chiral hinges modes, robust against disorders,
and propagate along the horizontal hinges in a cuboid geometry, shown in Fig.~\ref{fig1} (d)~\cite{SM}.
As type-II TSMs at top and bottom surfaces host opposite Chern numbers,
the corresponding chiral hinge modes show opposite chirality.
Note that chiral hinges states coexist with the gapped type-I TSMs in the bulk bandgap.
Although the $Z_{2}\times Z$ topology is built at finite $K_{x}$, where the PTRS is broken,
the topology is still preserved at $K_{x}=0$,
as the bulk polarization along stacking direction is independent on $K_{x}$.
In this situation, both the type-I SMs and chiral hinge modes are gapless and coexist in the bandgap~\cite{SM}.

\textit{Discussions and summary--}Finally, we discuss the possibility of experimental realization of our proposal.
The CrI$_{3}$, CrBr$_{3}$ and other honeycomb ferromagnets
with AFM interlayer coupling are promising candidates, and even the kagome ferromagnets.
There are already experimental observations of the topology of magnons in monolayer honeycomb ferromagnets, due to the DM interaction~\cite{LChen1,LChen2,FZhu},
which is necessary in our proposal.
The stacking type is not restricted in the AA-type. For example, the bilayer CrI$_{3}$
have two stable stacking, i.e., rhombohedral stacking with FM interlayer coupling and monoclinic stacking with AFM interlayer coupling.
The monoclinic stacking bilayer has an interlayer $\frac{1}{3}$-glide along the zigzag direction compared to AA stacking.
With the EPA term, we can also get the second-order topological phase~\cite{SM} when the two layers have an interlayer $\frac{1}{3}$-glide,
as the combined symmetry of $C_{2}1^{\prime}$ and
the glide plays as the effective mirror symmetry of $\mathcal{M}_{y}$ in AA stacking.
The $Z_{2}\times Z$ topology is independent on the stacking type, the chiral hinge modes are maintained in any stacking type.
In addition, the magnetic anisotropy terms, especially EPA term,
can be controlled by light, mechanical strain and electronic gating~\cite{DAfanasiev,KNakamura,YWang,IAVerzhbitskiy}
in 2D monolayer and van der Waals magnets.

In summary, we find the second-order topological phases of magnons in van der Waals ferromagnets with AFM interlayer coupling.
There are two different hinge modes coexisting arising from two different mechanisms.
Firstly, the AFM interlayer coupling realizes a magnonic Z$_{2}$ TI phase, protected by the PTRS.
In the presence of the EPA, the PTRS is broken, but the magnetic two-fold rotational symmetry gives rise to
rotation-graded topological polarization, which protects
a second-order topological phase.
Secondly, the introduced staggered AFM interlayer coupling brings a $Z_{2}\times Z$ topology, giving rise to type-II gapped TSMs carrying finite Chern numbers, thus supporting chiral hinge modes along horizontal directions.
Our work reveals the higher-order topological phase of magnons in 2D Van der Waals magnets, and paves an new way
for dissipationless magnonic devices using van der Waals magnets.

Y.-M. Li thanks Dr. Weinan Lin for his helpful discussions. This work is partly supported by MOST of China (Grants No. 2017YFA0303400).
 Y.-M. Li is supported by the startup funding from Xiamen University.

\begin{widetext}

\renewcommand\thefigure{S\arabic{figure}}
\renewcommand{\theequation}{S\arabic{equation}}

\section{Supplementary Materials}

\subsection{S1. Bosonic Bogoliubov-de Gennes Hamiltonian}

The spin interaction Hamiltonian is given by
\begin{equation}\label{eqS1}
  H=\sum_{l}H_{l}+\sum_{i,l}[J_{\perp}+(-1)^{l+1}\delta J_{\perp}]\mathbf{S}_{i}^{l}\cdot\mathbf{S}_{i}^{l+1},
\end{equation}
and the intralayer spin Hamiltonian
\begin{eqnarray}
  H_{l}=&-&J_{\parallel}\sum_{\langle ij\rangle}\mathbf{S}_{i}^{l}\cdot\mathbf{S}_{j}^{l}
  +D\sum_{\langle\langle ij\rangle\rangle}\nu_{ij}\hat{z}\cdot(\mathbf{S}_{i}^{l}\times\mathbf{S}_{j}^{l}) \nonumber \\
  &+&\sum_{i}[K_{x}(S_{i,x}^{l})^2-K_{z}(S_{i,z}^{l})^2]. \label{eqS2}
\end{eqnarray}
Due to interlayer antiferromagntic (AFM) order, we apply the Holstein-Primakoff transformation $S_{i,+}^{1}=\sqrt{2S}a_{i}^{1}$, $S_{i,-}^{1}=\sqrt{2S}a_{i}^{1\dagger}$, $S_{i,z}^{1}=S-a_{i}^{1\dagger}a_{i}^{1}$ for first layer
and $S_{i,-}^{2}=\sqrt{2S}a_{i}^{2}$, $S_{i,+}^{2}=\sqrt{2S}a_{i}^{2\dagger}$, $S_{i,z}^{2}=-S+a_{i}^{2\dagger}a_{i}^{2}$ for second layer,
we can get the  Hamiltonian up to the quadratic term
\begin{equation}\label{eqS3}
  H=\sum_{i,l}[h_{0}a_{i}^{l\dagger}a_{i}^{l}+(K_{x}a_{i}^{l}a_{i}^{l}+h_{\perp}^{l}a_{i}^{l}a_{i}^{l+1}+H.c.)]-\sum_{\langle ij\rangle,l}J_{\parallel}S(a_{i}^{l\dagger}a_{j}^{l}+H.c.)+\sum_{\langle\langle ij\rangle\rangle,l}(-1)^{l}DS(\nu_{ij}a_{i}^{l\dagger}a_{j}^{l}+H.c.),
\end{equation}
where $h_{0}=(2K_{z}+K_{x}+3J_{\parallel}+2J_{\perp})S$ and $h_{\perp}^{l}=[J_{\perp}-(-1)^{l}\delta J_{\perp}]S$, $H.c.$ denote the Hermitian conjugate.
Considering the AFM interlayer and FM intralayer  coupling, we apply the Fourier transformation,
\begin{equation*}
  a_{i}^{l}=\frac{1}{N}\sum_{\mathbf{k}}e^{i\mathbf{k}_{\parallel}\cdot\mathbf{r}_{i}^{l}}e^{ik_{z}r_{i,z}^{l}}b_{\alpha,\mathbf{k}}^{l}, \quad a_{i}^{l\dagger}=\frac{1}{N}\sum_{\mathbf{k}}e^{-i\mathbf{k}_{\parallel}\cdot\mathbf{r}_{i}^{l}}e^{-ik_{z}r_{i,z}^{l}}b_{\alpha,\mathbf{k}}^{l\dagger};\quad  l\in odd
\end{equation*}
\begin{equation*}
  a_{i}^{l}=\frac{1}{N}\sum_{\mathbf{k}}e^{i\mathbf{k}_{\parallel}\cdot\mathbf{r}_{i}^{l}}e^{-ik_{z}r_{i,z}^{l}}b_{\alpha,\mathbf{k}}^{l}, \quad a_{i}^{l\dagger}=\frac{1}{N}\sum_{\mathbf{k}}e^{-i\mathbf{k}_{\parallel}\cdot\mathbf{r}_{i}^{l}}e^{ik_{z}r_{i,z}^{l}}b_{\alpha,\mathbf{k}}^{l\dagger}.\quad  l\in even
\end{equation*}
$\mathbf{k}_{\parallel}=(k_{x},k_{y})$. Then we can get the Hamiltonian in momentum space $H=\sum_{\mathbf{k}}\Psi_{\mathbf{k}}^{\dagger}H_{\mathbf{k}}\Psi_{\mathbf{k}}$ with the Bosonic BdG Hamiltonian
\begin{equation}\label{eqS4}
  H_{\mathbf{k}}=\left(\begin{array}{cccc}
  h_{1}(\mathbf{k}) & 0 & K_{x} & h_{\perp,k_{z}} \\
  0 & h_{2}(\mathbf{k}) & h_{\perp,k_{z}} & K_{x} \\
  K_{x} & h_{\perp,k_{z}}^{\dagger} & h_{1}^{T}(-\mathbf{k}) & 0 \\
  h_{\perp,k_{z}}^{\dagger} & K_{x} & 0 &  h_{2}^{T}(-\mathbf{k})
  \end{array}\right).
\end{equation}
where the basis is $\Psi_{\mathbf{k}}=(b_{A,\mathbf{k}}^{1},b_{B,\mathbf{k}}^{1},b_{A,\mathbf{k}}^{2},b_{B,\mathbf{k}}^{2},
b_{A,-\mathbf{k}}^{1\dagger},b_{B,-\mathbf{k}}^{1\dagger},b_{A,-\mathbf{k}}^{2\dagger},b_{B,-\mathbf{k}}^{2\dagger})^{T}$.
Note that $-\mathbf{k}=(-k_{x},-k_{y},k_{z})$ appears in the subscript of basis and in hole part of the Hamiltonian.
$h_{1}(\mathbf{k})=h_{0}\sigma_{0}+h_{x}\sigma_{x}+h_{y}\sigma_{y}+h_{z}\sigma_{z}$ and $h_{2}(\mathbf{k})=h_{0}\sigma_{0}+h_{x}\sigma_{x}+h_{y}\sigma_{y}-h_{z}\sigma_{z}$. $\sigma_{0}$ is the $2\times2$ identity matrice
and $\sigma_{i=x,y,z}$ are the Pauli matrices.
$h_{\perp,k_{z}}=J_{\perp}(1+e^{-ik_{z}})+\delta J_{\perp}(1-e^{-ik_{z}})$.

\subsection{S2. Calculation of the Z$_{2}$ number}
We first focus on the bilayer case. The Hamiltonian host a inversion symmetry: $\mathcal{P}=\tau_{z}s_{0}\sigma_{x}$. The eigenvalues
at four time-reversal-invariant momentum (TRIM), i.e. $\Gamma$ and three $M$ points (shown in Fig.~S1 (a)), are $1$, $1$, $-1$, $1$, respectively.
So that $(-1)^{Z_{z}}=-1$ indicate $Z_{1}=1$ for bilayer system.

Now we focus on the 3D case. Four Z2 invariants are needed to form
the combination [$\nu_{0}$,($\nu_{1}$,$\nu_{2}$,$\nu_{3}$)] to describe the topological phase in 3D and distinguish the "strong" and "weak".
Here it is hard to directly calculate the four topological invariant as there is no inversion symmetry. We enlarge the unit cell along the $y$ direction
and the Brillouin zone (BZ) is reduced to a rectangle in the $k_{x}$-$k_{y}$ plane.
We adopt the method introduced by Rui Yu, et al.~\cite{RYu} to calculate the evolution lines of Wannier centers
to get the topological invariant indirectly. The effective BZ adopted is shown in Fig.~S1 (b) as the shaded region.
At $k_{z}=0$ and $k_{z}=\pi$, the Wannier centers both winding one time, indicating $\nu_{0}=0$, $\nu_{3}=1$.
at $k_{x}=\pi$ and $k_{y}=\pi$, the Wannier centers both winding zero times, $\nu_{1}=0$, $\nu_{2}=0$.
Thus the bulk system is in a "weak" TI phase.

\begin{figure}
  \centering
  \includegraphics[width=0.7\textwidth]{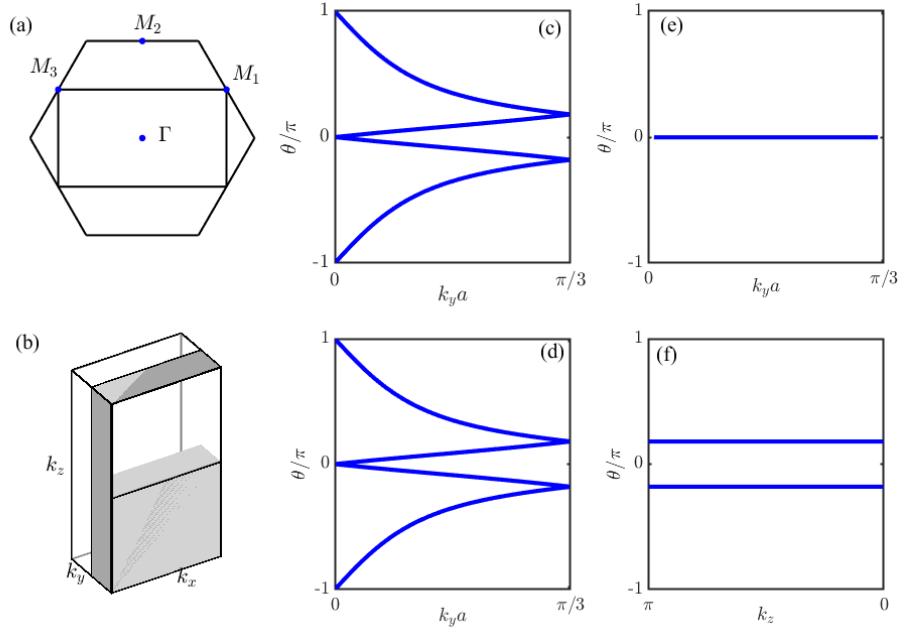}\\
  \caption{(a). The 2D BZ and the reduced BZ. (b) The reduced 3D BZ. The shaded region is the region for computing the Z$_{2}$ number using the Wilson loop method. (c-f). The Wannier center evolution. at the four shaded region in (b).}\label{figs1}
\end{figure}

\begin{figure}[b]
  \centering
  \includegraphics[width=0.5\textwidth]{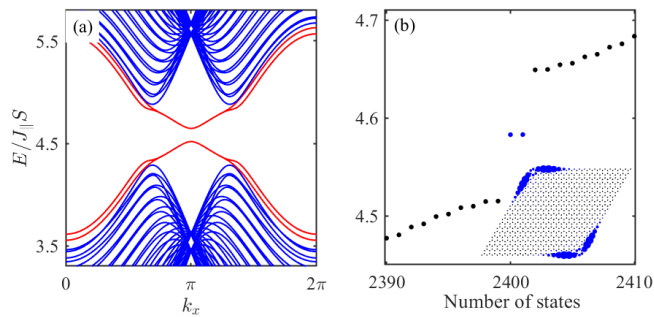}\\
  \caption{(a) Magnon bands of bilayer zigzag ribbons. The parameters are $J_{\perp}=0.3J_{\parallel}$, $D=0.05J_{\parallel}$, $K_{z}=0.2J_{\parallel}$,
  $K_{x}=J_{\parallel}$. In this and subsequent calculations, we adopted a compensated boundary.
  (b) Magnon eigenvalues in a parallelogram flake geometry. The inset shows the module square of wavefunction distribution of the corner modes.}\label{figs2}
\end{figure}

\subsection{S3. Corner Modes in bilayer}
The bilayer share a same Hamiltonian to Eq.~(3) in the main text just by setting $h_{0}=2K_{z}+K_{x}+3J_{\parallel}+J_{\perp}$
and $h_{\perp,k_{z}}=J_{\perp}$.
Then the bosonic Hamiltonian have an additional pseudo-inversion symmetry: $\mathcal{P}\tau_{z}H_{\mathbf{k}}\mathcal{P}=\tau_{z}H_{-\mathbf{k}}$
with $\mathcal{P}=\mathcal{P}^{-1}=\tau_{z}s_{0}\sigma_{x}$.
When $K_{x}=0$, the parity eigenvalues at $\Gamma$ and three M-points, i.e. the time-reversal invariant momentum (TRIM) points,
are $1$, $1$, $-1$, $1$, respectively, indicating $Z_{2}$ invariant $\nu=1$.
The bilayer system is thus topologically nontrivial and supports gapless edge states with two-components propagating oppositely at each edge.
The EPA breaks the pseudo-time-reversal symmetry and destroys the $Z_{2}$ topological phase and opens gap for zigzag-edged ribbon, as shown in Fig.~S2(a).
The same to the bulk case, the magnetic two-fold rotational symmetry $\mathcal{M}_{y}$ is preserved and rotation-graded topological invariant
along the line $k_{x}$ at $k_{y}=0$ is half-quantized, makes the system a second-order topological insulator.
The corner modes arise as a result, as shown in Fig.~S2(b) in a parallelogram flake.

\subsection{S4. The chiral Hinge modes at $K_{x}=0$}

In Fig.~S2 (a), we plot the magnon local density distribution of the chiral hinge modes. In (b), we plot the magnon bands along $k_{x}$-direction at $K_{x}=0$.
Without the EPA, there also exist the chiral hinge modes, coexisting with the gapless type-I TSMs, closing the gap.

\begin{figure}[t]
  \centering
  \includegraphics[width=0.5\textwidth]{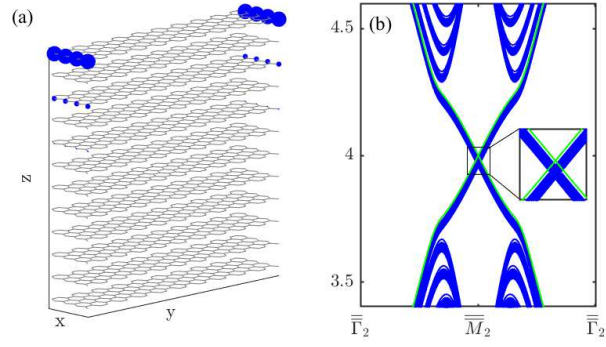}\\
  \caption{(a). Magnon local density distribution at $k_{x}=\pi$ at upper surface hinges. (b). The magnon bands along $k_{x}$-direction at $K_{x}=0$.
  The other parameters are the same to Fig.~3 (b) in the main text.}\label{figS3}
\end{figure}

\begin{figure}[b]
  \centering
  \includegraphics[width=0.8\textwidth]{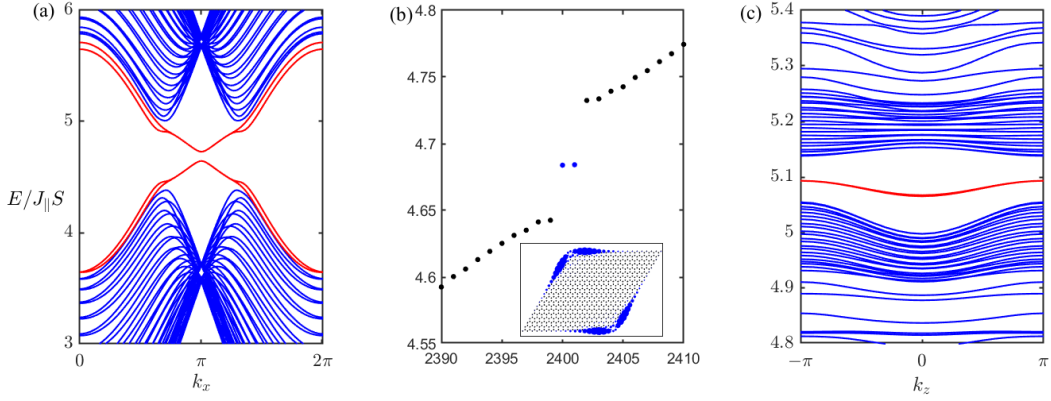}\\
  \caption{(a). The ribbon band of a bilayer honeycomb ferromagnetis with monoclinic stacking. The parameters are $J_{\perp}=0.2J_{\parallel}$,
  $D=0.05J_{\parallel}$, $K_{z}=0.2J_{\parallel}$, $K_{x}=J_{\parallel}$. (b). The corner modes in a parallelogram flake geometry. The inset
  shows the local density distribution of the down layer. The upper layers shares a similar distribution.
  (c). The magnon spectrum along $k_{z}$ direction with $\delta J_{\perp}=0.1J_{\parallel}$.}\label{figS4}
\end{figure}

\subsection{S6. Magnons in Monoclinic stacked honeycomb ferromagnets}

We first discuss the bilayer case to get an intuitive picture.
The monoclinic stacking of bilayer CrI$_{3}$ is a $\frac{1}{3}$-glide along the Zigzag direction compared with AA stacking.
Here we only consider the nearest interlayer AFM coupling.
The bosonic BdG Hamiltonian in $\mathbf{k}$-space is given by
\begin{equation}\label{eqS5}
  H_{\mathbf{k}}=\left(\begin{array}{cccc}
  h_{1}(\mathbf{k}) & 0 & K_{x} & h_{\perp,\mathbf{k}}^{0} \\
  0 & h_{2}(\mathbf{k}) & h_{\perp,-\mathbf{k}}^{0T} & K_{x} \\
  K_{x} & h_{\perp,-\mathbf{k}}^{0*} & h_{1}^{T}(-\mathbf{k}) & 0 \\
  h_{\perp,\mathbf{k}}^{0\dagger} & K_{x} & 0 &  h_{2}^{T}(-\mathbf{k})
  \end{array}\right),
\end{equation}
in the basis $\Psi_{\mathbf{k}}=(\psi_{\mathbf{k}},\psi_{-\mathbf{k}}^{\dagger})^{T}$ with
$\psi_{k}=(b_{A,\mathbf{k}}^{1},b_{B,\mathbf{k}}^{1},e^{i\frac{k_{x}}{\sqrt{3}}}b_{A,\mathbf{k}}^{2},e^{i\frac{k_{x}}{\sqrt{3}}}b_{B,\mathbf{k}}^{2})^{T}$.
Here $\mathbf{k}=(k_{x},k_{y})$,
\begin{equation*}
  h_{\perp,\mathbf{k}}^{0}=\left(\begin{array}{cc}
  J_{\perp} & J_{\perp} \\
  J_{\perp}e^{-i\sqrt{3}k_{x}} & J_{\perp} \end{array}
  \right)
\end{equation*}
For above bosonic Hamiltonian, it is hard to find the pseudo-time-reversal-symmetry (PTRS). But the magnon bands are still doubly degenerate when $K_{x}=0$.
This is because in this case, the bosonic Hamiltonian can be divided into two sectors and and EPA describes the coupling between the two sectors.
In two sectors, the interlayer coupling are the same with respect to a $k_{x}$-dependent phase factor. This factor coming from the interlayer glide.
For the AFM coupling, the obtained magnon bands are only dependent on the module of the coupling after Bogoliubov transformation.
So the magnon bands are doubly degenerate when $K_{x}=0$.
The EPA term couples the two sectors and will breaks the degeneracy.
In Fig.~S4 (a) and (b), we can see the EPA opens gap for the degenerate edge states and gives us second-order topological magnons, i.e., corner modes.

For the bulk with monoclinic materials, there should be 6 layers in a primitive cell to build a period along $z$-direction when considering the magnetic order.
Here to simplify our calculation, we only consider
a 2-layer period along $z$-direction. That is to say, the first and third layer are in AA-stacking. The interlayer interaction Hamiltonian
is
\begin{equation*}
  h_{\perp,\mathbf{k}}=\left(\begin{array}{cc}
  1 & 1 \\
  e^{-i\sqrt{3}k_{x}} & 1 \end{array}
  \right)\times [(J_{\perp}+\delta J_{\perp})+(J_{\perp}-\delta J_{\perp})e^{-ik_{z}}].
\end{equation*}
By considering a parallelogram pillar geometry, we can also get the hinge modes along the stacking direction, as shown in Fig.~S4 (c)
The bulk polarization along the stacking direction coming from the SSH configuration, independent on
the stacking type. The $Z_{2}\times Z$ topology is preserved and chiral hinge modes are expected to remain
in the monoclinic stacking.

\end{widetext}

\end{document}